\documentclass[pdflatex,sn-mathphys-num]{sn-jnl}

\usepackage{graphicx}%
\usepackage{multirow}%
\usepackage{amsmath,amssymb,amsfonts}%
\usepackage{amsthm}%
\usepackage{mathrsfs}%
\usepackage[title]{appendix}%
\usepackage{xcolor}%
\usepackage{textcomp}%
\usepackage{manyfoot}%
\usepackage{booktabs}%
\usepackage{algorithm}%
\usepackage{algorithmicx}%
\usepackage{algpseudocode}%
\usepackage{listings}%

\theoremstyle{thmstyleone}%
\theoremstyle{thmstyletwo}%

\theoremstyle{thmstylethree}%

\begin{document}

\title[Article Title]{Anisotropic compact star model with quadratic equation of state in paraboloidal spacetime}


\author[1]{\fnm{B. S.} \sur{Ratanpal}}\email{bharatratanpal@gmail.com}

\author*[1]{\fnm{Bhavesh} \sur{Suthar}}\email{bhaveshsuthar.math@gmail.com}

\affil[1]{\orgdiv{Department of Applied Mathematics, Faculty of Technology \& Engineering}, \orgname{The Maharaja Sayajirao University of Baroda}, \orgaddress{\city{Vadodara}, \postcode{390001}, \state{Gujarat}, \country{India}}}








\abstract{In this work, we report a new exact solution of Einstein’s field equations for static spherically symmetric anisotropic matter distributions on the background of paraboloidal spacetime by assuming a quadratic equation of state. The model parameters were found by matching the interior spacetime metric with the Schwarzschild exterior metric. The new exact solution is non-singular and meets all of the physical plausibility conditions for a realistic compact star. The detailed physical analysis of the model reveals that the gravitational potentials and matter variables are well-behaved throughout the distribution. The stability of the model has been analyzed under various conditions.}




\maketitle

\section{Introduction}\label{sec1}

In astronomy, the word 'compact object' refers to white dwarfs, neutron stars, and black holes. Stars are generally known to be isolated bodies bounded by self-gravity that emit energy from an internal source. Most of compact objects evolve at a point in stellar evolution when a star's internal radiation pressure from nuclear fusion is insufficient to balance its external gravitational force, causing the star to collapse under its own weight. It is well known that the model of a compact star may be generated through solving Einstein's field equations within the framework of general relativity. In general relativity and quantum gravity, these equations are used to investigate the distinct structures and features of stellar objects. They are also used to develop physical theories such as the light spectral shift and deflection for massive bodies, as well as the perihelion progression of planets in the solar system (\cite{Schwarzschild}, \cite{Will}, \cite{Clifton}). A number of approaches have been used for modifying general relativity theories, including perfect fluids, anisotropic matter distributions, charged fluid, higher-dimensional stars, and exotic matter configurations (\cite{Mathias}, \cite{Maurya2016}, \cite{Abdalla}, \cite{Bezerra}, \cite{Bhar2017}). These models also give intriguing insights into the interior geometry and behavior of nuclear matter, as predicting the maximum mass-radius ratio is dependent on model parameters as well as the equation of states (EoS), while taking into account isotropic and anisotropic fluid distributions. Einstein’s field equations are a system of highly non-linear second-order partial differential equations. To reduce such complexity in the context of stellar interiors, for example, the space-time metric can be assumed static and spherically symmetric. The latter converts the problem of second-order partial differential equations to ordinary second-order differential equations. Even though the equations remain strongly coupled, the problem is immensely simplified. However, depending on the global components of which matter distribution of fluid sphere is composed, that is, isotropic or anisotropic content with or without the presence of an electric charge, solving the ensuing equations is challenging. The simplest way to solve the equations analytically is to assume that the material composition is isotropic pr = pt (without considering electric charge) \cite{Tolman}. However, in order to provide a more realistic situation, local anisotropy must be included (\cite{MH2002}, \cite{MH2003}). Schwarzschild \cite{Schwarzschild} and Nordström \cite{Nordstrom} gave the first solutions to Einstein's field equations. Schwarzschild proposed a vacuum solution to investigate gravitationally non-rotating uncharged entities in spherical space-time with constant mass density. The Reissner-Nordström solution is a static solution to the Einstein-Maxwell field equations, representing the external gravitational field of a charged, non-rotating body. Aside from these solutions, Leijon \cite{Leijon} mentions several more well-known solutions to Einstein's field equations. These involve the Kerr–Newman solution to the spacetime associated with a charged and rotating body and the Kerr solution to the spacetime associated with a rotating body. The well-studied Vaidya-Tikekar model is analyzed in \cite{Sharma2006}, which shows that studies of EoS together with ansatz are equally important for a physically plausible model of super-dense compact objects.

The most significant aspect of modeling compact stellar objects is pressure anisotropy. At ultrahigh densities, the radial and tangential pressures within the stellar core may differ, and the pressure inside the stellar object may be anisotropic \cite{Ruderman1972}. Several researchers (\cite{Herrera1997}, \cite{Herrera1997-2}, \cite{MH2002a}, \cite{MH2002}, \cite{Chan2003}, \cite{HM2004}) have examined how local anisotropy impacts astrophysical objects and their origins. Pressure anisotropy can be caused by a solid core or the presence of type-3A superfluid \cite{HM2004}, phase transitions during gravitational collapse (\cite{Sokolov}, \cite{HN1989}), pion condensation (\cite{Sokolov}, \cite{HS1994}), slow rotation of a fluid \cite{HS1994}, viscosity \cite{I2010}, and strong electromagnetic fields (\cite{W1999}, \cite{MR2003}, \cite{U2004}). Karmakar et al. \cite{KM2007} observed that the existence of anisotropic pressures can cause a rise in redshifts, maximum compactness, and mass. The pressure anisotropy also affects the stability of gravitating compact objects \cite{GD2004}.

To describe the interior structure of the stellar system, researchers have employed several types of mathematical approaches to solve highly non-linear Einstein's field equations. The assumption of a relationship between radial pressure and energy density, $p=p\left(\rho\right)$, is known as the equation of state (EoS) method. The EoS method is very useful in understanding the physical behavior of compact objects. This method is helpful for modeling already discovered compact objects as well as providing a feasible solution to theoretical problems in cosmology and relativistic astrophysics. Numerous authors have proposed different EoS models to find realistic solutions to Einstein's field equations for stellar objects. Sharma and Maharaj \cite{SM2007} employed a linear equation of state to construct relativistic compact models that are compatible with observational data. Ngubelanga and Maharaj \cite{NM2015} investigated physically feasible relativistic compact star models with a linear equation of state in isotropic coordinates. Alberto et al. \cite{MS2023} utilized a linear equation of state to generate a model of a charged anisotropic compact star. Patel et al. \cite{PR2023} employed a linear equation of state to explain the interior of an anisotropic compact star. Thirukkanesh and Ragel \cite{TR2012} and Takisa and Maharaj \cite{TM2013} used a polytropic EoS to produce the results for relativistic stars. Prasad et al. \cite{PJ2021} utilized the Chaplygin equation to develop a model of an anisotropic compact star.

A simple generalisation of the linear relation between the energy density and radial pressure is a quadratic equation of state, $p_{r} = \alpha \rho^{2}\left(r\right)+\beta \rho\left(r\right)-\gamma$. This quadratic form, which is more flexible than the linear equation of state, $p_{r} = \beta \rho\left(r\right)-\gamma$, can better describe complicated interactions in high-density regimes typical of compact stellar objects. The quadratic term $\alpha \rho^{2}\left(r\right)$ provides a more realistic representation of a matter's behavior under extreme conditions. Using a quadratic EoS is an important advancement since the complexity of the model may be significantly raised by including the nonlinearity of radial pressure in terms of energy density. Various forms of quadratic EoS are frequently employed in numerous research works. Sharma and Ratanpal \cite{SR2013} developed a model of stars with a quadratic EoS using the Finch-Skea geometry. Feroze and Siddiqui \cite{FS2011} used a quadratic equation of state to develop a model for charged anisotropic matter distributions. Ngubelanga et al. \cite{NM2015-1} employed a quadratic equation of state to establish a charged anisotropic compact star model in isotropic coordinates. Sunzu and Thomas \cite{ST2018} developed a neutral model using a quadratic equation of state with a vanishing anisotropic property. Using a quadratic equation of state, Malaver and Kasmaei \cite{MD2020} created a relativistic stellar model for charged anisotropic matter distribution. Thirukkanesh et al. \cite{TB2021} developed a favorable model by combining the Vaidya-Tikekar gravitational potential with the quadratic equation of state. Bhar et al. \cite{BS2016} constructed a model of an anisotropic compact star using a quadratic equation of state and a physically reasonable metric potential. Kumar et al. \cite{KK2024} proposed a model for a compact star using a quadratic equation of state and Buchdahl metric potential.

Inspired by all of these above-mentioned works, in this work, we obtain a new model of a singularity-free anisotropic compact star by assuming a quadratic equation of state, coupled with a suitable form of the metric potential. It is found that our solution that describes the interior of the star is stable, physically feasible, non-singular, continuous and maintains hydrostatic equilibrium in the interior of the star. The workflow of this paper is as follows: In Section \ref{sec2}, paraboloidal spacetime and the Einstein's field equations for anisotropic matter distribution have been discussed. By considering the quadratic equation of state along with the ansatz $e^{\lambda} = 1 + \frac{r^{2}}{R^{2}}$, the system is solved in section \ref{sec3}. In Section \ref{sec4}, the interior solution is matched with the Schwarzschild exterior line element over the boundary. A detailed discussion on the physical viability and stability of our model is presented in Section \ref{sec5}, and finally, in Section \ref{sec6}, we conclude our work.

\section{The spacetime metric}\label{sec2}
The Cartesian equation for a 3-paraboloid immersed in a 4-dimensional Euclidean space is
\begin{equation} \label{e1}
    x^{2}+y^{2}+z^{2}=2\omega R,
\end{equation}
where $\omega$ = constants yields a sphere and $x, y$ and $z$ = constants yield three-paraboloids.
Using the parametrization
\begin{eqnarray*}
	x &=& r \sin{\theta} \cos{\phi},\\
	y &=& r \sin{\theta} \sin{\phi},\\
	z &=& r \cos{\theta},\\
	\omega &=& \frac{r^2}{2R},
\end{eqnarray*}
the Euclidean metric
\begin{equation}\label{e2}
	d\sigma^2 = dx^2 + dy^2 +dz^2 +d\omega^2
\end{equation}
becomes
\begin{equation}\label{e3}
	ds^2 =\left(1+\frac{r^2}{R^2}\right) dr^2 -r^2(d\theta^2 +\sin^2{\theta} d\phi^{2}).
\end{equation}
Consider the interior spacetime of a static spherically symmetric stellar configuration in canonical coordinates as
\begin{equation}\label{IM}
	ds^2 = e^{\nu\left(r\right)} dt^2-e^{\lambda\left(r\right)} dr^2 -r^2(d\theta^2 +\sin^2{\theta}  d\phi^2)
\end{equation}
where
\begin{equation} \label{elam}
    e^{\lambda\left(r\right)}=1+\frac{r^2}{R^2}.
\end{equation}
The constant $ \frac{1}{R^2} $  can be identified with the curvature parameter. Following Maharaj and Maartens \cite{MM1989}, the energy-momentum tensor for an anisotropic matter distribution is taken as
\begin{equation} \label{EMT}
    T_{ij}=\left(\rho+p\right)u_{i}u_{j}-pg_{ij}+\sqrt{3}S\left[C_{i}C_{j}-\frac{1}{3}\left(u_{i}u_{j}-g_{ij}\right)\right],
\end{equation}
where $\rho$, $p$, and $u_{i}$ are the energy density, isotropic pressure, and fluid 4-velocity, respectively. $S=S(r)$ represents the magnitude of anisotropy, whereas $C_{i} = \left(0, e^{-\lambda^{2}}, 0, 0\right)$ is a radially directed vector. The nonvanishing components of the energy-momentum tensor are
\begin{equation} \label{EMT1}
    T_{0}^{0}=\rho, \hspace{0.2in} T_{1}^{1}=-\left(p+\frac{2S}{\sqrt{3}}\right), \hspace{0.2in} T_{2}^{2}=T_{3}^{3}=-\left(p-\frac{S}{\sqrt{3}}\right),
\end{equation}
implying that the radial and tangential pressures will have the forms
\begin{equation}
    p_{r}=-T_{1}^{1}=p+\frac{2S}{\sqrt{3}}
\end{equation}
and
\begin{equation}
    p_{t}=-T_{2}^{2}=p-\frac{S}{\sqrt{3}}
\end{equation}
respectively. Therefore, the magnitude of the anisotropy is obtained as
\begin{equation}
    p_{r}-p_{t}=\sqrt{3}S.
\end{equation}
The system of Einstein's field equations corresponds to the spacetime metric (\ref{IM}) and the energy-momentum tensor (\ref{EMT}) obtained as (in relativistic units with $G = c = 1$)
\begin{equation} \label{rho}
    8\pi\rho=\frac{1-e^{-\lambda}}{r^{2}}+\frac{e^{-\lambda}\lambda^{'}}{r},
\end{equation}
\begin{equation} \label{pr}
    8\pi p_{r}=\frac{e^{-\lambda}-1}{r^{2}}+\frac{e^{-\lambda}\nu^{'}}{r},
\end{equation}
\begin{equation}
    8\pi p_{t}=\frac{e^{-\lambda}}{4}\left[2\nu^{''}+\left(\nu^{'}-\lambda^{'}\right)\left(\nu^{'}+\frac{2}{r}\right)\right],
\end{equation}
where primes (') denote differentiation with respect to the radial coordinate r. Substituting  $e^{\lambda}=1+\frac{r^2}{R^2}$ into equations (\ref{rho}) and (\ref{pr}) yields 

\begin{equation} \label{rho1}
    8\pi \rho =\frac{r^{2}+3R^{2}}{\left(r^{2}+R^{2}\right)^{2}},
\end{equation}

\begin{equation} \label{pr1}
    8\pi p_{r} =\frac{-r+\nu^{'}R^{2}}{r\left(r^{2}+R^{2}\right)}.
\end{equation}

\section{Quadratic equation of state} \label{sec3}
The assumption of the equation of state is essential to develop a physically acceptable stellar model that connects pressure with the density of the star. Most of the earlier works were centered on implementing a linear equation of state of the form $p = \alpha \rho$, where $\alpha$ is a constant. We employ the quadratic equation of state of the form
\begin{equation} \label{eos}
    p_{r}=\alpha\left(r\right) \rho^{2}+\beta\left(r\right) \rho,
\end{equation}
with $\alpha\left(r\right)=A\left(1-\frac{r^{2}}{R^{2}}\right)$ and $\beta\left(r\right)=B\left(1-\frac{r^{2}}{R^{2}}\right)$,
where $A$ and $B$ are constants. Maharaj and Takisa \cite{MM2012}, Bhar et al. \cite{Bhar2017}, Christopher et al. \cite{CJ2024}, and Takisa et al. \cite{MM2019} all used the quadratic EoS to model compact stars within the framework of general relativity.\\
Substituting the equations (\ref{rho1}) and (\ref{pr1}) into the equation (\ref{eos}), we get

\begin{equation}
    \nu^{'}=\frac{1}{R^{2}}\left[r-\frac{A r\left(r^{2}-R^{2}\right)\left(r^{2}+3R^{2}\right)^{2}}{R^{2}\left(r^{2}+R^{2}\right)^{3}}-\frac{B r \left(r^{2}-R^{2}\right) \left(r^{2}+3R^{2}\right)}{R^{2}\left(r^{2}+R^{2}\right)}\right],
\end{equation}
which is integrable and yields
\small
\begin{equation} \label{enu1}
    e^{\nu}=C \left(r^{2}+R^{2}\right)^{2B-\frac{A}{R^{2}}} \exp{\left[\frac{-Br^{4}}{4R^{4}}-\frac{2AR^{2}}{\left(r^{2}+R^{2}\right)^{2}}-\frac{2A}{r^{2}+R^{2}}-\frac{r^{2}\left(A+B R^{2}-R^{2}\right)}{2R^{4}}\right]},
\end{equation}
\normalsize
where $C$ is a constant of integration. Therefore, the interior spacetime (\ref{IM}) takes the form
\small
\begin{multline} \label{Imetric1}
ds^2 = C \left(r^{2}+R^{2}\right)^{2B-\frac{A}{R^{2}}} \exp{\left[\frac{-Br^{4}}{4R^{4}}-\frac{2AR^{2}}{\left(r^{2}+R^{2}\right)^{2}}-\frac{2A}{r^{2}+R^{2}}-\frac{r^{2}\left(A+B R^{2}-R^{2}\right)}{2R^{4}}\right]} dt^2 \\ - \left(1+\frac{r^{2}}{R^{2}}\right) dr^2 -r^2(d\theta^2 +\sin^2{\theta}  d\phi^2).
\end{multline}
\normalsize

\section{Matching Condition} \label{sec4}
For a physically reasonable relativistic distribution of matter, the interior spacetime metric should continuously match with the exterior metric. We use the Israel-Darmois junction conditions (\cite{I1966}, \cite{D1927}) to smoothly match the interior spacetime metric (\ref{Imetric1}) with the Schwarzschild exterior metric
\begin{equation}
    ds^{2}=\left(1-\frac{2M}{R}\right)dt^{2}-\left(1-\frac{2M}{R}\right)^{-1}dr^{2}-r^2(d\theta^2 +\sin^2{\theta}  d\phi^2)
\end{equation}
over the boundary of the star $r = R$. The continuity of the metric coefficients $e^{\nu}$ and $e^{\lambda}$ across the boundary $r = R$, i.e.,
\begin{equation}
    e^{-\lambda\left(r\right)}|_{r=R}=e^{\nu\left(r\right)}|_{r=R}=1-\frac{2M}{R},
\end{equation}
is called the first fundamental form and the second fundamental form of the junction condition is that the radial pressure should vanish at the surface, i.e.,
\begin{equation}
    p_{r}\left(R\right)=0.
\end{equation}
Using these boundary conditions, we get
\begin{equation} \label{MR}
    R=4M,
\end{equation}
\begin{equation} \label{C}
    C=2^{\frac{A}{R^2}-2B-1} R^{\frac{2A}{R^2}-4B} \exp{\left(\frac{2A}{R^2}+\frac{3B}{4}-\frac{1}{2}\right)}
\end{equation}
and
\begin{equation}
    \frac{B}{A}=-\frac{r^2+3R^2}{\left(r^2+R^2\right)^2}
\end{equation}
where $M$ is the total mass enclosed within the boundary surface $R$. If the radius $R$ is known, Equation (\ref{MR}) may be used to calculate the total mass $M$ of the star, and vice versa. Substituting equation (\ref{C}) into equation (\ref{enu1}), we get
\large
\begin{equation}
    e^{\nu}=2^{\frac{A}{R^{2}}-2B-1} R^{\frac{2A}{R^2}-4B} \left(r^{2}+R^{2}\right)^{\frac{2BR^{2}-A}{R^2}} \exp{F},
\end{equation}
\normalsize
where
\begin{equation*}
    F=\frac{\left(R^{2}-r^{2}\right)\left[2A\left(r^4-r^2 R^2-4R^4\right)+\left(r^2 +R^2\right)^{2} \left(Br^{2}+3BR^{2}-2R^2\right)\right]}{4R^{4}\left(r^2 +R^2\right)^2}.
\end{equation*}
Subsequently, the expressions for energy density, radial pressure and tangential pressure can be given as
\begin{equation}
    \rho=\frac{r^2+3R^2}{\left(r^2+R^2\right)^2},
\end{equation}
\begin{equation}
    p_{r}=\frac{\left(3R^4-2r^2R^2-r^4\right)\left[A\left(r^2+3R^2\right)+B\left(r^2+R^2\right)^2\right]}{\left(r^2+R^2\right)^4},
\end{equation}
and
\small
\begin{equation}
    p_{t}=\frac{A^2r^2\left(r^2-R^2\right)^2 \left(r^2+3R^2\right)^4 +2A\left(r^2+R^2\right)^2\left(r^2+3R^2\right)F_{1} +\left(r^2+R^2\right)^4 F_{2}}{4R^6\left(r^2+R^2\right)^7},
\end{equation}
\normalsize
where
\begin{equation*}
    F_{1}=Br^{10}+\left(4B-1\right)r^8R^2-2\left(2+B\right)r^6 R^4 + \left(1-2B\right)r^4 R^6+9\left(B-2\right)r^2 R^8+6R^{10}
\end{equation*}
and
\begin{multline*}
    F_{2}=B^2r^{10}+2B\left(2B-1\right)r^8R^2+\left[1-2B\left(B+6\right)\right]r^6R^4-2\left[B\left(6B+7\right)-2\right]r^4R^6 \\+\left[3+B\left(9B-16\right)r^2R^8+12BR^{10}\right].
\end{multline*}
The implementation of the preceding section’s requirements, which guarantee the solution is viable, allows us to calculate the range of constants. In order to validate our model for a known star, we consider compact star \textbf{4U 1820–30}, whose radius $(R)$ is $9.1 km$ \cite{GR2013}. Using these values, we obtain the constant of integration $C$ from equation (\ref{C}). The range for the parameters $\alpha$ and $\beta$ can be established by utilizing the physical acceptability conditions described in section 5.

\section{Physical Analysis} \label{sec5}
Delgaty and Lake \cite{DL1998} examined 127 solutions, and only 16 passed the elementary test for physical relevance, with just 9 qualifying for the decreasing sound speed from center to surface of the star. Therefore, every physically acceptable stellar interior solution should satisfy the following physical acceptability conditions (\cite{MF2015}, \cite{K1988}, \cite{K1972}, \cite{B1979}). In the subsections below, we examined the model's stability and acceptability through different conditions:
\subsection{Regularity of the metric potentials}
Any realistic stellar model should be free from physical and geometric singularities. To verify this, the gravitational potentials $\lambda$ and $\nu$ must be continuous, finite, and regular. The metric functions must meet the conditions $e^{\lambda}=1$ and $e^{\nu}=$ constant at centre of the star $r=0$.\\
In our model,
\begin{equation*}
    e^{\lambda\left(r=0\right)}=1
\end{equation*}
and
\begin{equation*}
    e^{\nu\left(r=0\right)}=2^{\frac{A}{R^2}-2B-1}\exp{\left[\frac{3B}{4}-\frac{2A}{R^2}-\frac{1}{2}\right]},
\end{equation*}
which are constants, and $\left(e^{\lambda}\right)^{'}_{r=0}=\left(e^{\nu}\right)^{'}_{r=0}=0$, which imply that the gravitational potentials are regular at the origin.

\subsection{Nature of density and pressure}
\begin{figure}
\centering
    \includegraphics[height=.25\textheight]{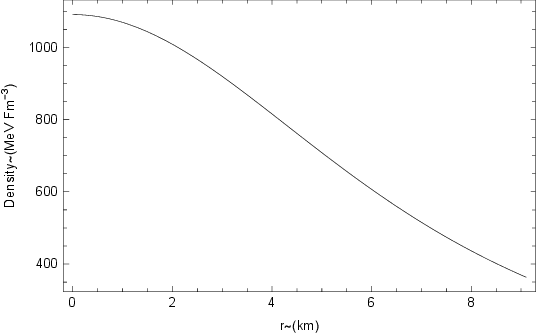}
    \caption{Variation of energy density ($\rho$) with respect to radial coordinate $(r)$}
    \label{fig:Figure. 1}
\end{figure}

\begin{figure}
\centering
    \includegraphics[height=.25\textheight]{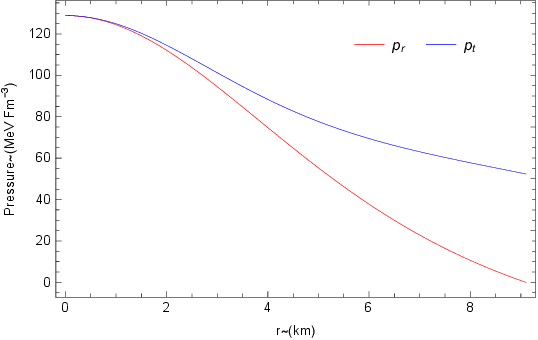}
    \caption{Variation of pressures ($p_{r}$, and $p_{t}$) with respect to radial coordinate $(r)$}
    \label{fig:Figure. 2}
\end{figure}

The regularity criteria for exact solutions are also investigated in the context of anisotropic pressure. Anisotropic fluids allow pressures to rise rapidly in various spatial directions. For spherical symmetry, spatial directions are specified in terms of radial pressure and tangential pressure. Radial pressure $\left(p_{r}\right)$ and tangential pressure $\left(p_{t}\right)$ should be equal at the centre of the star and decrease monotonically from the origin to the surface of the star (\cite{M2018}, \cite{SR2016}). The energy density $\left(\rho\right)$ of stellar objects should not be negative and decrease monotonically across the distribution.\\
i.e., $\rho\left(r\right)$, $p_{r}\left(r\right)$, $p_{t}\left(r\right)\geq0$ and $\frac{d\rho}{dr}, \frac{dp_{r}}{dr}, \frac{dp_{t}}{dr}<0,$ for $0\leq r \leq R$.\\
Since
\begin{equation*}
    \rho\left(0\right)=\frac{3}{R^2},
\end{equation*}
\begin{equation*}
    p_{r}\left(0\right)=\frac{9A+3BR^{2}}{R^4}
\end{equation*}
and
\begin{equation*}
    p_{t}\left(0\right)=\frac{9A+3BR^{2}}{R^4},
\end{equation*}
which implies the energy density and pressures are positive and regular at the origin. Figures \ref{fig:Figure. 1} and \ref{fig:Figure. 2} illustrate the variation of density and pressure from the centre to the surface of the star, respectively. It can be shown that energy density and pressure are monotonically decreasing functions of r.

\subsection{Pressure Anisotropy}
\begin{figure}
\centering
    \includegraphics[height=.25\textheight]{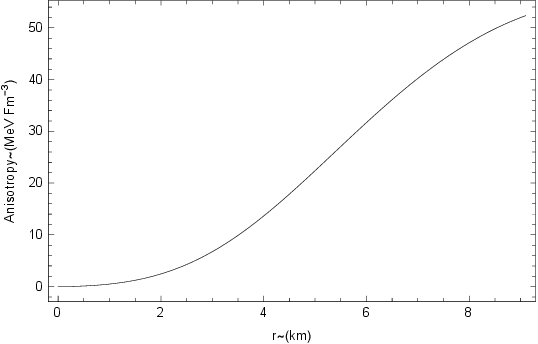}
    \caption{Measure of Anisotropy ($\Delta$) against radial coordinate $(r)$}
    \label{fig:Figure. 3}
\end{figure}

The anisotropic factor $(\Delta)$ is the difference between tangential and radial pressure, which will be repulsive or directed outwards if $p_{t} > p_{r}$ and attractive or directed inward when $p_{t} < p_{r}$. The property of the pressure anisotropy $(\Delta)$ is that it should be zero at the center of a compact star, indicating that the radial and tangential pressures are equal at the center of the star; in other words, at the center, the pressure appears isotropic in nature. A physically valid model requires $p_{t}(r=R) > 0$ and $p_{r}(r=R) = 0$; therefore, $\Delta (r = R)$ is always positive. At the same time, it should be positive within the stellar interior, as a positive anisotropic factor makes the system more stable, and it helps to construct more compact objects \cite{GM1994}. In our models, $\Delta(r=0) = 0$ and increases positively within the stellar interiors, as illustrated in Figure \ref{fig:Figure. 3}.

\subsection{Causality Condition}
\begin{figure}
\centering
    \includegraphics[height=.25\textheight]{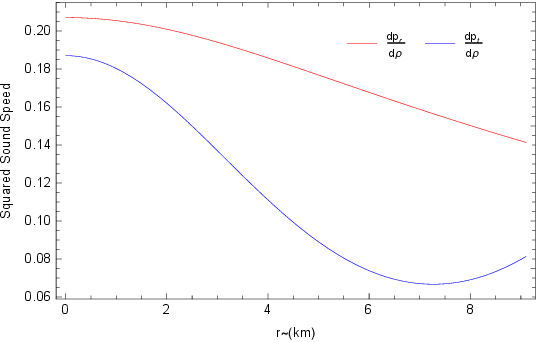}
    \caption{Squared Sound Speed against radial coordinate ($r$)}
    \label{fig:Figure. 4}
\end{figure}

For a physically viable anisotropic stellar structure, the subliminal speed of sound must be less than the speed of light. When dealing with anisotropic fluids, pressure waves propagate along the main directions, namely radial and tangential directions. The radial and tangential velocity of sound are derived from the expressions
\begin{equation*}
    {v^{2}_{r}}=\frac{dp_{r}}{d\rho},    
    {v^{2}_{t}}=\frac{dp_{t}}{d\rho}.
\end{equation*}
To achieve a physically acceptable model, both sound speeds ($v_{r}$ and $v_{t}$) must be restricted by the speed of light, which is known as the causality condition. The causality condition states that both the radial and transverse velocity of sound should be less than one, i.e., $0\leq{v^{2}_{r}},{v^{2}_{t}}\leq1$ (\cite{H1992}, \cite{AH2007}). The preservation/non-preservation of the causality condition has significant consequences for matter distribution throughout the structure. This is because it is linked to the behaviour of the energy-momentum tensor, which defines material composition. Due to the complexity of the expression, it is quite difficult to verify the causality condition analytically. We analyzed this condition using a graphical representation. Figure \ref{fig:Figure. 4} clearly shows that both $v^{2}_{r}$ and $v^{2}_{t}$ are within the acceptable range.

\subsection{Energy Conditions}
\begin{figure}
\centering
    \includegraphics[height=.25\textheight]{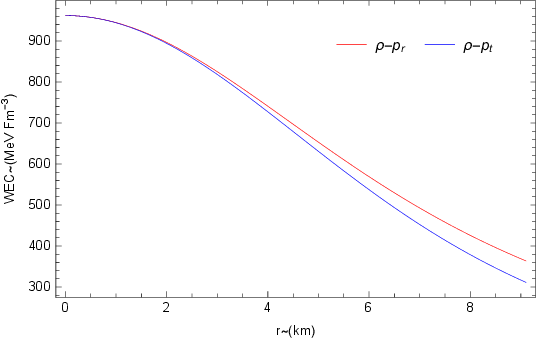}
    \caption{Weak Energy Condition against radial coordinate ($r$)}
    \label{fig:Figure. 5}
\end{figure}

\begin{figure}
\centering
    \includegraphics[height=.25\textheight]{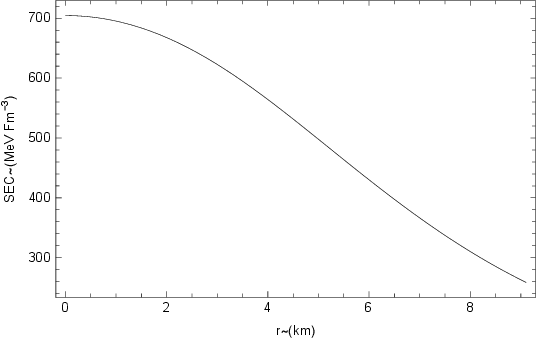}
    \caption{Strong Energy Condition against radial coordinate ($r$)}
    \label{fig:Figure. 6}
\end{figure}

Since compact stars are thought to be made up of a large number of material components. Although the components of the matter distribution are well-known, describing the exact form of the energy-momentum tensor may be quite difficult. The term "energy conditions" refers to a set of physical constraints that could be utilized to identify the presence of ordinary or exotic matter inside the stellar configuration. These conditions are defined as (\cite{P1993}, \cite{V1995}).
\begin{equation} \label{NEC}
    \text{Null Energy Condition (NEC): }\rho\ge0,
\end{equation}
\begin{equation}
    \text{Weak Energy Condition (WEC): }\rho-p_{r}\ge0, \rho-p_{t}\ge0,
\end{equation}
\begin{equation} \label{SEC}
    \text{Strong Energy Condition (SEC): }\rho-p_{r}-2p_{t}\ge0.
\end{equation}
If the anisotropic compact star meets the constraints listed above, the energy-momentum tensor is positive throughout the configuration. According to Maurya and Tello-Ortiz \cite{MT2019}, violations of energy conditions have been described as being caused only by unphysical stress energy tensors. Figures \ref{fig:Figure. 5} and \ref{fig:Figure. 6} display the energy condition graphs, and our model satisfies all the above energy conditions.

\subsection{Adiabatic Index}
\begin{figure}
\centering
    \includegraphics[height=.25\textheight]{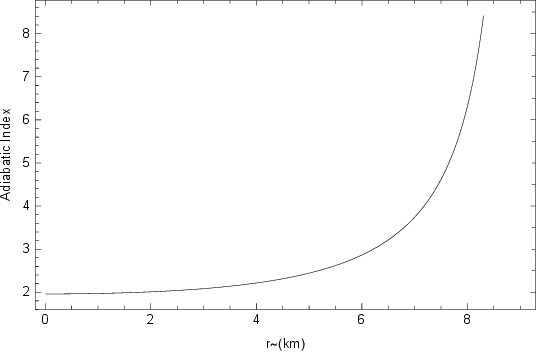}
    \caption{Adiabatic Index ($\Gamma$) against radial coordinate $(r)$}
    \label{fig:Figure. 7}
\end{figure}

The term adiabatic index refers to the stiffness of the equation of state for a density, and it also describes the stability of both relativistic and non-relativistic compact stars. According to Chan et al. \cite{CH1993}, the adiabatic index $(\Gamma)$ is defined as
\begin{equation*}
    \Gamma=\left(\frac{\rho+p_{r}}{p_{r}}\right)\frac{dp_{r}}{d\rho}.
\end{equation*}
This quantity, known as Bondi's criteria \cite{B1964}, determines whether a gravitational collapse occurs under small radial perturbations. According to this criterion, any relativistic fluid system will experience gravitational collapse if $\Gamma=\frac{4}{3}$ and be catastrophic if $\Gamma<\frac{4}{3}$. To overcome this issue, Moustakidis \cite{M2017} proposed a more strict condition $\Gamma\ge\Gamma_{crit}$, where $\Gamma_{crit}=\frac{4}{3}+\frac{19}{21}u$ is the critical value for the adiabatic index with $u$ being the compactness factor. Figure \ref{fig:Figure. 7} depicts a graph of the adiabatic index $(\Gamma)$. The figure indicates that the adiabatic index $(\Gamma)$ is a monotonically increasing function of $r$, with $\Gamma>\frac{4}{3}$ throughout the stellar configuration, confirming stability.

\subsection{Redshift}
\begin{figure}
\centering
    \includegraphics[height=.25\textheight]{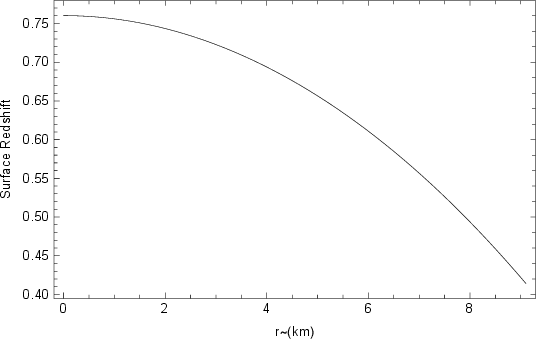}
    \caption{Surface Redfhift against radial coordinate $(r)$}
    \label{fig:Figure. 8}
\end{figure}

The redshift plays an evolving role in understanding the strong physical interaction between particles within the compact object and its equation of state (EoS). According to Straumann \cite{S1984} and Buchdahl \cite{B1959}, the maximum redshift can’t exceed 2 for isotropic stellar configurations in the absence of a cosmological constant.  Bowers and Liang \cite{BL1974} demonstrated that pressure anisotropy can surpass this upper bound.  Karmakar et al. \cite{KM2007} and Barraco et al. \cite{BH2003} proposed that the redshift for the anisotropic stellar model could be 3.84.  Furthermore, Böhmer and Harko \cite{BH2006} demonstrated that the value of redshift could rise up to $z\leq5$.  For a relativistic star, the redshift $z=\sqrt{e^{-\nu\left(r\right)}}-1$ should decrease towards the boundary and remain finite throughout the distribution.  In our model, redshift decreases with r and $z<0.76$, as seen in Figure \ref{fig:Figure. 8}.

\section{Conclusion}\label{sec6}
In this work, we obtained an exact analytic solution to Einstein's field equations for a static spherically symmetric anisotropic matter distribution within the framework of general relativity. The singularity-free solution was derived by assuming a quadratic equation of state in the context of paraboloidal spacetime. We matched our interior solution to the Schwarzschild exterior solution at the boundary to discover the values of the different constants. We investigated the physical viability of our stellar model, considering the radius of the compact star \textbf{4U 1820-30} as an input parameter (we assumed $R = 9.1 km$). We observe that the metric coefficients are free from singularity. Figure \ref{fig:Figure. 1} shows that the energy density ($\rho$) is a monotonically decreasing function of the radial coordinate $r$, with a maximum value in the center of the star. Figure \ref{fig:Figure. 2} shows that the radial pressure ($p_r$) and tangential pressure ($p_t$) are positive within the stellar interior and monotonically decreasing towards the boundary. As previously stated, the anisotropy factor $\Delta = p_t-p_r$ is zero in the center of the star and increases radially outward. Since $\Delta > 0$, the anisotropic force is repulsive in nature, which is favourable for the stability of our model. In addition, at the centre of the star, $\Delta$ vanishes (as seen in figure \ref{fig:Figure. 3}); that makes the model more realistic. The model also meets the causality condition, which states that the square of sound velocity is positive and less than one everywhere within a compact star, as seen in figure \ref{fig:Figure. 4}. Figures \ref{fig:Figure. 5} and \ref{fig:Figure. 6} clearly demonstrates that the model meets the null, weak, and strong energy conditions. The stability of the stellar configuration has been checked through the adiabatic index ($\Gamma$), as seen in figure \ref{fig:Figure. 7}. The redshift is decreasing from the center of the star to its boundary, as seen in Figure \ref{fig:Figure. 8}. A key aspect of the model is that the exact solution derived is simple, which is found in few solutions. It has been demonstrated that the compact object satisfies all physical requirements for being physically viable and stable. We've just shown the physical analysis for one compact star here, but it can be extended to a wider range of known pulsars. As a result, the model developed here can play an important role in the description of the internal structure of compact stars within the framework of general relativity.

\backmatter

\bmhead{Acknowledgements}
For providing research facilities, the authors are grateful to the Inter-University Centre for Astronomy and Astrophysics (IUCAA), Pune, India.

\bibliography{manuscript}


\begin{thebibliography}{72}
\ifx \bisbn   \undefined \def \bisbn  #1{ISBN #1}\fi
\ifx \binits  \undefined \def \binits#1{#1}\fi
\ifx \bauthor  \undefined \def \bauthor#1{#1}\fi
\ifx \batitle  \undefined \def \batitle#1{#1}\fi
\ifx \bjtitle  \undefined \def \bjtitle#1{#1}\fi
\ifx \bvolume  \undefined \def \bvolume#1{\textbf{#1}}\fi
\ifx \byear  \undefined \def \byear#1{#1}\fi
\ifx \bissue  \undefined \def \bissue#1{#1}\fi
\ifx \bfpage  \undefined \def \bfpage#1{#1}\fi
\ifx \blpage  \undefined \def \blpage #1{#1}\fi
\ifx \burl  \undefined \def \burl#1{\textsf{#1}}\fi
\ifx \doiurl  \undefined \def \doiurl#1{\url{https://doi.org/#1}}\fi
\ifx \betal  \undefined \def \betal{\textit{et al.}}\fi
\ifx \binstitute  \undefined \def \binstitute#1{#1}\fi
\ifx \binstitutionaled  \undefined \def \binstitutionaled#1{#1}\fi
\ifx \bctitle  \undefined \def \bctitle#1{#1}\fi
\ifx \beditor  \undefined \def \beditor#1{#1}\fi
\ifx \bpublisher  \undefined \def \bpublisher#1{#1}\fi
\ifx \bbtitle  \undefined \def \bbtitle#1{#1}\fi
\ifx \bedition  \undefined \def \bedition#1{#1}\fi
\ifx \bseriesno  \undefined \def \bseriesno#1{#1}\fi
\ifx \blocation  \undefined \def \blocation#1{#1}\fi
\ifx \bsertitle  \undefined \def \bsertitle#1{#1}\fi
\ifx \bsnm \undefined \def \bsnm#1{#1}\fi
\ifx \bsuffix \undefined \def \bsuffix#1{#1}\fi
\ifx \bparticle \undefined \def \bparticle#1{#1}\fi
\ifx \barticle \undefined \def \barticle#1{#1}\fi
\bibcommenthead
\ifx \bconfdate \undefined \def \bconfdate #1{#1}\fi
\ifx \botherref \undefined \def \botherref #1{#1}\fi
\ifx \url \undefined \def \url#1{\textsf{#1}}\fi
\ifx \bchapter \undefined \def \bchapter#1{#1}\fi
\ifx \bbook \undefined \def \bbook#1{#1}\fi
\ifx \bcomment \undefined \def \bcomment#1{#1}\fi
\ifx \oauthor \undefined \def \oauthor#1{#1}\fi
\ifx \citeauthoryear \undefined \def \citeauthoryear#1{#1}\fi
\ifx \endbibitem  \undefined \def \endbibitem {}\fi
\ifx \bconflocation  \undefined \def \bconflocation#1{#1}\fi
\ifx \arxivurl  \undefined \def \arxivurl#1{\textsf{#1}}\fi
\csname PreBibitemsHook\endcsname

\bibitem[\protect\citeauthoryear{Schwarzschild}{1916}]{Schwarzschild}
\begin{barticle}
\bauthor{\bsnm{Schwarzschild}, \binits{K.}}:
\batitle{On the gravitational field of a mass point according to einstein's theory}.
\bjtitle{Sitzungsber. Preuss. Akad. Wiss. Berlin (Math. Phys. )}
\bvolume{1916},
\bfpage{189}--\blpage{196}
(\byear{1916})
\end{barticle}
\endbibitem

\bibitem[\protect\citeauthoryear{Will}{2006}]{Will}
\begin{barticle}
\bauthor{\bsnm{Will}, \binits{C.M.}}:
\batitle{The confrontation between general relativity and experiment}.
\bjtitle{Living Rev. Relat.}
\bvolume{9}(\bissue{1}),
\bfpage{3}
(\byear{2006})
\end{barticle}
\endbibitem

\bibitem[\protect\citeauthoryear{Clifton et~al.}{2012}]{Clifton}
\begin{barticle}
\bauthor{\bsnm{Clifton}, \binits{T.}},
\bauthor{\bsnm{Ferreira}, \binits{P.G.}},
\bauthor{\bsnm{Padilla}, \binits{A.}},
\bauthor{\bsnm{Skordis}, \binits{C.}}:
\batitle{Modified gravity and cosmology}.
\bjtitle{Phys. Rep.}
\bvolume{513},
\bfpage{1}--\blpage{3}
(\byear{2012})
\end{barticle}
\endbibitem

\bibitem[\protect\citeauthoryear{Mathias et~al.}{2021}]{Mathias}
\begin{barticle}
\bauthor{\bsnm{Mathias}, \binits{A.K.}},
\bauthor{\bsnm{Maharaj}, \binits{S.D.}},
\bauthor{\bsnm{Sunzu}, \binits{J.M.}},
\bauthor{\bsnm{Mkenyeleye}, \binits{J.M.}}:
\batitle{Charged anisotropic models via embedding}.
\bjtitle{Pramana– J. Phys.}
\bvolume{95},
\bfpage{178}
(\byear{2021})
\end{barticle}
\endbibitem

\bibitem[\protect\citeauthoryear{Maurya et~al.}{2016}]{Maurya2016}
\begin{barticle}
\bauthor{\bsnm{Maurya}, \binits{S.K.}},
\bauthor{\bsnm{Gupta}, \binits{Y.K.}},
\bauthor{\bsnm{Ray}, \binits{S.}},
\bauthor{\bsnm{Deb}, \binits{D.}}:
\batitle{Generalised model for anisotropic compact stars}.
\bjtitle{Eur. Phys. J. C}
\bvolume{76}(\bissue{12}),
\bfpage{693}
(\byear{2016})
\end{barticle}
\endbibitem

\bibitem[\protect\citeauthoryear{Abdalla et~al.}{2021}]{Abdalla}
\begin{barticle}
\bauthor{\bsnm{Abdalla}, \binits{A.T.}},
\bauthor{\bsnm{Sunzu}, \binits{J.M.}},
\bauthor{\bsnm{Mkenyeleye}, \binits{J.M.}}:
\batitle{Generalised charged anisotropic quark star models}.
\bjtitle{Pramana– J. Phys.}
\bvolume{95},
\bfpage{86}
(\byear{2021})
\end{barticle}
\endbibitem

\bibitem[\protect\citeauthoryear{Bezerra~de Mello}{2006}]{Bezerra}
\begin{barticle}
\bauthor{\bsnm{Mello}, \binits{E.R.}}:
\batitle{Vacuum polarization effects on flat branes due to a global monopole}.
\bjtitle{Phys. Rev. D}
\bvolume{73}(\bissue{10}),
\bfpage{105015}
(\byear{2006})
\end{barticle}
\endbibitem

\bibitem[\protect\citeauthoryear{Bhar et~al.}{2017}]{Bhar2017}
\begin{barticle}
\bauthor{\bsnm{Bhar}, \binits{P.}},
\bauthor{\bsnm{Singh}, \binits{K.N.}},
\bauthor{\bsnm{Pant}, \binits{N.}}:
\batitle{Compact star modeling with quadratic equation of state in tolman vii space–time}.
\bjtitle{Ind. J. Phys.}
\bvolume{91}(\bissue{6}),
\bfpage{701}--\blpage{709}
(\byear{2017})
\end{barticle}
\endbibitem

\bibitem[\protect\citeauthoryear{Tolman}{1939}]{Tolman}
\begin{barticle}
\bauthor{\bsnm{Tolman}, \binits{R.C.}}:
\batitle{Static solutions of einstein's field equations for spheres of fluid}.
\bjtitle{Phys. Rev.}
\bvolume{55}(\bissue{4}),
\bfpage{364}--\blpage{373}
(\byear{1939})
\end{barticle}
\endbibitem

\bibitem[\protect\citeauthoryear{Mak and Harko}{2002}]{MH2002}
\begin{barticle}
\bauthor{\bsnm{Mak}, \binits{M.K.}},
\bauthor{\bsnm{Harko}, \binits{T.}}:
\batitle{An exact anisotropic quark star model}.
\bjtitle{Chin. J. Astron. Astrophys.}
\bvolume{2}(\bissue{3}),
\bfpage{248}
(\byear{2002})
\end{barticle}
\endbibitem

\bibitem[\protect\citeauthoryear{Mak and Harko}{2003}]{MH2003}
\begin{barticle}
\bauthor{\bsnm{Mak}, \binits{M.K.}},
\bauthor{\bsnm{Harko}, \binits{T.}}:
\batitle{Anisotropic stars in general relativity}.
\bjtitle{Proc. R. Soc. Lond. A}
\bvolume{459}(\bissue{2030}),
\bfpage{393}--\blpage{408}
(\byear{2003})
\end{barticle}
\endbibitem

\bibitem[\protect\citeauthoryear{Nordstr{\"o}m}{1918}]{Nordstrom}
\begin{barticle}
\bauthor{\bsnm{Nordstr{\"o}m}, \binits{G.}}:
\batitle{On the energy of the gravitation field in einstein's theory}.
\bjtitle{Koninklijke Nederlandsche Akademie van Wetenschappen Proc.}
\bvolume{20},
\bfpage{1238}--\blpage{1245}
(\byear{1918})
\end{barticle}
\endbibitem

\bibitem[\protect\citeauthoryear{Leijon}{2012}]{Leijon}
\begin{botherref}
\oauthor{\bsnm{Leijon}, \binits{R.}}:
The Einstein Field Equations : on semi-Riemannian manifolds, and the Schwarzschild solution
(2012)
\end{botherref}
\endbibitem

\bibitem[\protect\citeauthoryear{Sharma et~al.}{2006}]{Sharma2006}
\begin{barticle}
\bauthor{\bsnm{Sharma}, \binits{R.}},
\bauthor{\bsnm{Karmakar}, \binits{S.}},
\bauthor{\bsnm{Mukherjee}, \binits{S.}}:
\batitle{Maximum mass of a class of cold compact stars}.
\bjtitle{Int. J. Mod. Phys. D}
\bvolume{15}(\bissue{03}),
\bfpage{405}--\blpage{418}
(\byear{2006})
\end{barticle}
\endbibitem

\bibitem[\protect\citeauthoryear{Ruderman}{1972}]{Ruderman1972}
\begin{barticle}
\bauthor{\bsnm{Ruderman}, \binits{M.}}:
\batitle{Pulsars: Structure and dynamics}.
\bjtitle{Ann. Rev. Astron. Astrophys.}
\bvolume{10},
\bfpage{427}--\blpage{476}
(\byear{1972})
\end{barticle}
\endbibitem

\bibitem[\protect\citeauthoryear{Herrera and Santos}{1997a}]{Herrera1997}
\begin{barticle}
\bauthor{\bsnm{Herrera}, \binits{L.}},
\bauthor{\bsnm{Santos}, \binits{N.O.}}:
\batitle{Local anisotropy in self-gravitating systems}.
\bjtitle{Phys. Rep.}
\bvolume{286}(\bissue{2}),
\bfpage{53}--\blpage{130}
(\byear{1997})
\end{barticle}
\endbibitem

\bibitem[\protect\citeauthoryear{Herrera and Santos}{1997b}]{Herrera1997-2}
\begin{barticle}
\bauthor{\bsnm{Herrera}, \binits{L.}},
\bauthor{\bsnm{Santos}, \binits{N.O.}}:
\batitle{Thermal evolution of compact objects and relaxation time}.
\bjtitle{R. Astron. Soc.}
\bvolume{287}(\bissue{1}),
\bfpage{161}--\blpage{164}
(\byear{1997})
\end{barticle}
\endbibitem

\bibitem[\protect\citeauthoryear{Mak and Harko}{2002}]{MH2002a}
\begin{barticle}
\bauthor{\bsnm{Mak}, \binits{M.K.}},
\bauthor{\bsnm{Harko}, \binits{T.}}:
\batitle{New method for generating general solution of abel differential equation}.
\bjtitle{Comp. \& Math. with Appl.}
\bvolume{43}(\bissue{1}),
\bfpage{91}--\blpage{94}
(\byear{2002})
\end{barticle}
\endbibitem

\bibitem[\protect\citeauthoryear{Chan et~al.}{2003}]{Chan2003}
\begin{barticle}
\bauthor{\bsnm{Chan}, \binits{R.}},
\bauthor{\bsnm{DA~Silva}, \binits{M.F.A.}},
\bauthor{\bsnm{Da~Rocha}, \binits{J.F.V.}}:
\batitle{Gravitational collapse of selfsimilar and shear free fluid with heat flow}.
\bjtitle{Int. J. Mod. Phys.s D}
\bvolume{12}(\bissue{3}),
\bfpage{347}--\blpage{368}
(\byear{2003})
\end{barticle}
\endbibitem

\bibitem[\protect\citeauthoryear{Harko and Mak}{2004}]{HM2004}
\begin{barticle}
\bauthor{\bsnm{Harko}, \binits{T.}},
\bauthor{\bsnm{Mak}, \binits{M.K.}}:
\batitle{Anisotropy in bianchi-type brane cosmologies}.
\bjtitle{Class. Quant. Gravit.}
\bvolume{21}(\bissue{6}),
\bfpage{1489}
(\byear{2004})
\end{barticle}
\endbibitem

\bibitem[\protect\citeauthoryear{Sokolov}{1980}]{Sokolov}
\begin{barticle}
\bauthor{\bsnm{Sokolov}, \binits{A.I.}}:
\batitle{Phase transitions in a superfluid neutron liquid}.
\bjtitle{Soviet J. Exper. Theor. Phys.}
\bvolume{52},
\bfpage{575}
(\byear{1980})
\end{barticle}
\endbibitem

\bibitem[\protect\citeauthoryear{Herrera and Nunez}{1989}]{HN1989}
\begin{barticle}
\bauthor{\bsnm{Herrera}, \binits{L.}},
\bauthor{\bsnm{Nunez}, \binits{L.}}:
\batitle{Modeling 'hydrodynamic phase transitions' in a radiating spherically symmetric distribution of matter}.
\bjtitle{Astrophys. J.}
\bvolume{339},
\bfpage{339}--\blpage{353}
(\byear{1989})
\end{barticle}
\endbibitem

\bibitem[\protect\citeauthoryear{Herrera and Santos}{1995}]{HS1994}
\begin{barticle}
\bauthor{\bsnm{Herrera}, \binits{L.}},
\bauthor{\bsnm{Santos}, \binits{N.}}:
\batitle{Jeans mass for anisotropic matter}.
\bjtitle{Astrophys. J.}
\bvolume{438},
\bfpage{308}--\blpage{313}
(\byear{1995})
\end{barticle}
\endbibitem

\bibitem[\protect\citeauthoryear{Ivanov}{2010}]{I2010}
\begin{barticle}
\bauthor{\bsnm{Ivanov}, \binits{B.V.}}:
\batitle{The importance of anisotropy for relativistic fluids with spherical symmetry}.
\bjtitle{Int. J. Theor. Phys.}
\bvolume{49}(\bissue{6}),
\bfpage{1236}--\blpage{1243}
(\byear{2010})
\end{barticle}
\endbibitem

\bibitem[\protect\citeauthoryear{Weber}{1999}]{W1999}
\begin{barticle}
\bauthor{\bsnm{Weber}, \binits{F.}}:
\batitle{Quark matter in neutron stars}.
\bjtitle{J. Phys. G: Nuc. Part. Phys.}
\bvolume{25}(\bissue{9}),
\bfpage{195}
(\byear{1999})
\end{barticle}
\endbibitem

\bibitem[\protect\citeauthoryear{P{\'e}rez~Mart{\'\i}nez et~al.}{2003}]{MR2003}
\begin{barticle}
\bauthor{\bsnm{P{\'e}rez~Mart{\'\i}nez}, \binits{A.}},
\bauthor{\bsnm{P{\'e}rez~Rojas}, \binits{H.}},
\bauthor{\bsnm{Mosquera~Cuesta}, \binits{H.J.}}:
\batitle{Magnetic collapse of a neutron gas: Can magnetars indeed be formed?}
\bjtitle{Eur. Phys.l J. C}
\bvolume{29}(\bissue{1}),
\bfpage{111}--\blpage{123}
(\byear{2003})
\end{barticle}
\endbibitem

\bibitem[\protect\citeauthoryear{Usov}{2004}]{U2004}
\begin{barticle}
\bauthor{\bsnm{Usov}, \binits{V.V.}}:
\batitle{Electric fields at the quark surface of strange stars in the color-flavor locked phase}.
\bjtitle{Phys. Rev. D}
\bvolume{70},
\bfpage{067301}
(\byear{2004})
\end{barticle}
\endbibitem

\bibitem[\protect\citeauthoryear{Karmakar et~al.}{2007}]{KM2007}
\begin{barticle}
\bauthor{\bsnm{Karmakar}, \binits{S.}},
\bauthor{\bsnm{Mukherjee}, \binits{S.}},
\bauthor{\bsnm{Sharma}, \binits{R.}},
\bauthor{\bsnm{Maharaj}, \binits{S.D.}}:
\batitle{The role of pressure anisotropy on the maximum mass of cold compact stars}.
\bjtitle{Pramana - J. Phys.}
\bvolume{68},
\bfpage{881}--\blpage{889}
(\byear{2007})
\end{barticle}
\endbibitem

\bibitem[\protect\citeauthoryear{Gleiser and Dev}{2004}]{GD2004}
\begin{barticle}
\bauthor{\bsnm{Gleiser}, \binits{M.}},
\bauthor{\bsnm{Dev}, \binits{K.}}:
\batitle{Anistropic stars: Exact solutions and stability}.
\bjtitle{Int. J. Mod. Phys. D}
\bvolume{13}(\bissue{7}),
\bfpage{1389}--\blpage{1397}
(\byear{2004})
\end{barticle}
\endbibitem

\bibitem[\protect\citeauthoryear{Sharma and Maharaj}{2007}]{SM2007}
\begin{barticle}
\bauthor{\bsnm{Sharma}, \binits{R.}},
\bauthor{\bsnm{Maharaj}, \binits{S.D.}}:
\batitle{A class of relativistic stars with a linear equation of state}.
\bjtitle{Mon. Not. R. Astron. Soc.}
\bvolume{375}(\bissue{4}),
\bfpage{1265}--\blpage{1268}
(\byear{2007})
\end{barticle}
\endbibitem

\bibitem[\protect\citeauthoryear{Ngubelanga and Maharaj}{2015}]{NM2015}
\begin{barticle}
\bauthor{\bsnm{Ngubelanga}, \binits{S.}},
\bauthor{\bsnm{Maharaj}, \binits{S.D.}}:
\batitle{Relativistic stars with polytropic equation of state}.
\bjtitle{Eur. Phys. J. Plus}
\bvolume{130},
\bfpage{211}
(\byear{2015})
\end{barticle}
\endbibitem

\bibitem[\protect\citeauthoryear{Mathias et~al.}{2023}]{MS2023}
\begin{barticle}
\bauthor{\bsnm{Mathias}, \binits{A.}},
\bauthor{\bsnm{Sunzu}, \binits{J.}},
\bauthor{\bsnm{Maharaj}, \binits{S.D.}},
\bauthor{\bsnm{Mkenyeleye}, \binits{J.}}:
\batitle{Charged anisotropic model with embedding and a linear equation of state}.
\bjtitle{Pramana - J. Phys.}
\bvolume{97},
\bfpage{29}
(\byear{2023})
\end{barticle}
\endbibitem

\bibitem[\protect\citeauthoryear{Patel et~al.}{2023}]{PR2023}
\begin{barticle}
\bauthor{\bsnm{Patel}, \binits{R.}},
\bauthor{\bsnm{Ratanpal}, \binits{B.S.}},
\bauthor{\bsnm{Pandya}, \binits{D.M.}}:
\batitle{New charged anisotropic solution on paraboloidal spacetime}.
\bjtitle{Astrophys. Space Sci.}
\bvolume{368},
\bfpage{58}
(\byear{2023})
\end{barticle}
\endbibitem

\bibitem[\protect\citeauthoryear{Thirukkanesh and Ragel}{2012}]{TR2012}
\begin{barticle}
\bauthor{\bsnm{Thirukkanesh}, \binits{S.}},
\bauthor{\bsnm{Ragel}, \binits{F.}}:
\batitle{Exact anisotropic sphere with polytropic equation of state}.
\bjtitle{Pramana - J. Phys.}
\bvolume{78}(\bissue{5}),
\bfpage{687}--\blpage{696}
(\byear{2012})
\end{barticle}
\endbibitem

\bibitem[\protect\citeauthoryear{Takisa and Maharaj}{2013}]{TM2013}
\begin{barticle}
\bauthor{\bsnm{Takisa}, \binits{P.M.}},
\bauthor{\bsnm{Maharaj}, \binits{S.D.}}:
\batitle{Some charged polytropic models}.
\bjtitle{Gen. Relat. Gravit.}
\bvolume{45},
\bfpage{1951}--\blpage{1969}
(\byear{2013})
\end{barticle}
\endbibitem

\bibitem[\protect\citeauthoryear{Prasad et~al.}{2021}]{PJ2021}
\begin{barticle}
\bauthor{\bsnm{Prasad}, \binits{A.K.}},
\bauthor{\bsnm{Kumar}, \binits{J.}},
\bauthor{\bsnm{Sarkar}, \binits{A.}}:
\batitle{Behavior of anisotropic fluids with chaplygin equation of state in buchdahl spacetime}.
\bjtitle{Gen. Relat. Gravit.}
\bvolume{53}(\bissue{12}),
\bfpage{108}
(\byear{2021})
\end{barticle}
\endbibitem

\bibitem[\protect\citeauthoryear{Sharma and Ratanpal}{2013}]{SR2013}
\begin{barticle}
\bauthor{\bsnm{Sharma}, \binits{R.}},
\bauthor{\bsnm{Ratanpal}, \binits{B.S.}}:
\batitle{Relativistic stellar model admitting a quadratic equation of state}.
\bjtitle{Int. J. Mod. Phys. D}
\bvolume{22}(\bissue{13}),
\bfpage{1350074}
(\byear{2013})
\end{barticle}
\endbibitem

\bibitem[\protect\citeauthoryear{Feroze and Siddiqui}{2011}]{FS2011}
\begin{barticle}
\bauthor{\bsnm{Feroze}, \binits{T.}},
\bauthor{\bsnm{Siddiqui}, \binits{A.A.}}:
\batitle{Charged anisotropic matter with quadratic equation of state}.
\bjtitle{Gen. Relat. Gravit.}
\bvolume{43},
\bfpage{1025}--\blpage{1035}
(\byear{2011})
\end{barticle}
\endbibitem

\bibitem[\protect\citeauthoryear{Ngubelanga et~al.}{2015}]{NM2015-1}
\begin{barticle}
\bauthor{\bsnm{Ngubelanga}, \binits{S.A.}},
\bauthor{\bsnm{Maharaj}, \binits{S.D.}},
\bauthor{\bsnm{Ray}, \binits{S.}}:
\batitle{Compact stars with quadratic equation of state}.
\bjtitle{Astrophys. Space Sci.}
\bvolume{357}(\bissue{1}),
\bfpage{74}
(\byear{2015})
\end{barticle}
\endbibitem

\bibitem[\protect\citeauthoryear{Sunzu and Thomas}{2018}]{ST2018}
\begin{barticle}
\bauthor{\bsnm{Sunzu}, \binits{J.M.}},
\bauthor{\bsnm{Thomas}, \binits{M.}}:
\batitle{Newstellar models generated using a quadratic equation of state}.
\bjtitle{Pramana - J. Phys.}
\bvolume{91},
\bfpage{75}
(\byear{2018})
\end{barticle}
\endbibitem

\bibitem[\protect\citeauthoryear{Malaver and Daei~Kasmaei}{2020}]{MD2020}
\begin{barticle}
\bauthor{\bsnm{Malaver}, \binits{M.}},
\bauthor{\bsnm{Daei~Kasmaei}, \binits{H.}}:
\batitle{Relativistic stellar models with quadratic equation of state}.
\bjtitle{Int. J. Math. Mod. Comp.}
\bvolume{10}(\bissue{2}),
\bfpage{111}--\blpage{124}
(\byear{2020})
\end{barticle}
\endbibitem

\bibitem[\protect\citeauthoryear{Thirukkanesh et~al.}{2021}]{TB2021}
\begin{barticle}
\bauthor{\bsnm{Thirukkanesh}, \binits{S.}},
\bauthor{\bsnm{Bogadi}, \binits{R.S.}},
\bauthor{\bsnm{Govender}, \binits{M.}},
\bauthor{\bsnm{Moyo}, \binits{S.}}:
\batitle{Stability and improved physical characteristics of relativistic compact objects arising from the quadratic term in $p_r = \alpha \rho ^2 + \beta \rho - \gamma $}.
\bjtitle{Eur. Phys. J. C}
\bvolume{81},
\bfpage{62}
(\byear{2021})
\end{barticle}
\endbibitem

\bibitem[\protect\citeauthoryear{Bhar et~al.}{2016}]{BS2016}
\begin{barticle}
\bauthor{\bsnm{Bhar}, \binits{P.}},
\bauthor{\bsnm{Singh}, \binits{K.N.}},
\bauthor{\bsnm{Pant}, \binits{N.}}:
\batitle{Compact stellar models obeying quadratic equation of state}.
\bjtitle{Astrophys. Space Sci.}
\bvolume{361},
\bfpage{343}
(\byear{2016})
\end{barticle}
\endbibitem

\bibitem[\protect\citeauthoryear{Kumar et~al.}{2024}]{KK2024}
\begin{barticle}
\bauthor{\bsnm{Kumar}, \binits{M.}},
\bauthor{\bsnm{Kumar}, \binits{J.}},
\bauthor{\bsnm{Bharti}, \binits{P.}},
\bauthor{\bsnm{Prasad}, \binits{A.K.}}:
\batitle{Exploring the physics of relativistic compact stars: an anisotropic model with quadratic equation of state in buchdahl geometry}.
\bjtitle{Astrophys. Space Sci.}
\bvolume{369}(\bissue{9}),
\bfpage{97}
(\byear{2024})
\end{barticle}
\endbibitem

\bibitem[\protect\citeauthoryear{Maharaj and Maartens}{1989}]{MM1989}
\begin{barticle}
\bauthor{\bsnm{Maharaj}, \binits{S.D.}},
\bauthor{\bsnm{Maartens}, \binits{R.}}:
\batitle{Anisotropic spheres with uniform energy density in general relativity}.
\bjtitle{Gen. Relat. Gravit.}
\bvolume{21},
\bfpage{899}--\blpage{905}
(\byear{1989})
\end{barticle}
\endbibitem

\bibitem[\protect\citeauthoryear{Maharaj and Takisa}{2012}]{MM2012}
\begin{barticle}
\bauthor{\bsnm{Maharaj}, \binits{S.D.}},
\bauthor{\bsnm{Takisa}, \binits{M.P.}}:
\batitle{Regular models with quadratic equation of state}.
\bjtitle{Gen. Relat. Gravit.}
\bvolume{44},
\bfpage{1419}--\blpage{1432}
(\byear{2012})
\end{barticle}
\endbibitem

\bibitem[\protect\citeauthoryear{Christopher et~al.}{2024}]{CJ2024}
\begin{barticle}
\bauthor{\bsnm{Christopher}, \binits{J.}},
\bauthor{\bsnm{Jape}, \binits{J.W.}},
\bauthor{\bsnm{Sunzu}, \binits{J.M.}}:
\batitle{Charged anisotropic conformal star model with a quadratic equation of state}.
\bjtitle{Int. J. Mod. Phys. D}
\bvolume{33}(\bissue{05n06}),
\bfpage{2450022}
(\byear{2024})
\end{barticle}
\endbibitem

\bibitem[\protect\citeauthoryear{Takisa et~al.}{2019}]{MM2019}
\begin{barticle}
\bauthor{\bsnm{Takisa}, \binits{M.P.}},
\bauthor{\bsnm{Maharaj}, \binits{S.D.}},
\bauthor{\bsnm{Mulangu}, \binits{C.}}:
\batitle{Compact relativistic star with quadratic envelope}.
\bjtitle{Pramana - J. Phys.}
\bvolume{92},
\bfpage{40}
(\byear{2019})
\end{barticle}
\endbibitem

\bibitem[\protect\citeauthoryear{Israel}{1966}]{I1966}
\begin{barticle}
\bauthor{\bsnm{Israel}, \binits{W.}}:
\batitle{Singular hypersurfaces and thin shells in general relativity}.
\bjtitle{Nuovo Cimento B (1965-1970)}
\bvolume{44},
\bfpage{1}--\blpage{14}
(\byear{1966})
\end{barticle}
\endbibitem

\bibitem[\protect\citeauthoryear{Darmois}{1927}]{D1927}
\begin{barticle}
\bauthor{\bsnm{Darmois}, \binits{G.}}:
\batitle{The equations of einsteinian gravitation}.
\bjtitle{Memorial Math. Sci.}
\bvolume{25},
\bfpage{58}
(\byear{1927})
\end{barticle}
\endbibitem

\bibitem[\protect\citeauthoryear{Gangopadhyay et~al.}{2013}]{GR2013}
\begin{barticle}
\bauthor{\bsnm{Gangopadhyay}, \binits{T.}},
\bauthor{\bsnm{Ray}, \binits{S.}},
\bauthor{\bsnm{Li}, \binits{X.-D.}},
\bauthor{\bsnm{Dey}, \binits{J.}},
\bauthor{\bsnm{Dey}, \binits{M.}}:
\batitle{Strange star equation of state fits the refined mass measurement of 12 pulsars and predicts their radii}.
\bjtitle{Mon. Not. R. Astron. Soc.}
\bvolume{431}(\bissue{4}),
\bfpage{3216}--\blpage{3221}
(\byear{2013})
\end{barticle}
\endbibitem

\bibitem[\protect\citeauthoryear{Delgaty and Lake}{1998}]{DL1998}
\begin{barticle}
\bauthor{\bsnm{Delgaty}, \binits{M.S.R.}},
\bauthor{\bsnm{Lake}, \binits{K.}}:
\batitle{Physical acceptability of isolated, static, spherically symmetric, perfect fluid solutions of einstein's equations}.
\bjtitle{Comp. Phys. Commun.}
\bvolume{115}(\bissue{2}),
\bfpage{395}--\blpage{415}
(\byear{1998})
\end{barticle}
\endbibitem

\bibitem[\protect\citeauthoryear{Murad and Fatema}{2015}]{MF2015}
\begin{barticle}
\bauthor{\bsnm{Murad}, \binits{M.H.}},
\bauthor{\bsnm{Fatema}, \binits{S.}}:
\batitle{Some new wyman–leibovitz–adler type static relativistic charged anisotropic fluid spheres compatible to self-bound stellar modeling}.
\bjtitle{Eur. Phys. J. C}
\bvolume{75},
\bfpage{533}
(\byear{2015})
\end{barticle}
\endbibitem

\bibitem[\protect\citeauthoryear{Knusten}{1988}]{K1988}
\begin{barticle}
\bauthor{\bsnm{Knusten}, \binits{H.}}:
\batitle{Report on the physical characteristics of vaidya-tikekar's exact relativistic model for a superdense star}.
\bjtitle{Astron. Nachr.}
\bvolume{309},
\bfpage{263}--\blpage{265}
(\byear{1988})
\end{barticle}
\endbibitem

\bibitem[\protect\citeauthoryear{Kuchowicz}{1972}]{K1972}
\begin{barticle}
\bauthor{\bsnm{Kuchowicz}, \binits{B.}}:
\batitle{Differential conditions for physically meaningful fluid spheres in general relativity}.
\bjtitle{Phys. Lett. A}
\bvolume{38}(\bissue{5}),
\bfpage{369}--\blpage{370}
(\byear{1972})
\end{barticle}
\endbibitem

\bibitem[\protect\citeauthoryear{Buchdahl}{1979}]{B1979}
\begin{barticle}
\bauthor{\bsnm{Buchdahl}, \binits{H.A.}}:
\batitle{Regular general relativistic charged fluid spheres}.
\bjtitle{Acta Phys. Pol.}
\bvolume{10}(\bissue{8}),
\bfpage{673}--\blpage{685}
(\byear{1979})
\end{barticle}
\endbibitem

\bibitem[\protect\citeauthoryear{Murad}{2018}]{M2018}
\begin{barticle}
\bauthor{\bsnm{Murad}, \binits{M.H.}}:
\batitle{Some families of relativistic anisotropic compact stellar models embedded in pseudo-euclidean space $e^5$: an algorithm}.
\bjtitle{Eur. Phys. J. C}
\bvolume{78},
\bfpage{285}
(\byear{2018})
\end{barticle}
\endbibitem

\bibitem[\protect\citeauthoryear{Shee et~al.}{2016}]{SR2016}
\begin{barticle}
\bauthor{\bsnm{Shee}, \binits{D.}},
\bauthor{\bsnm{Rahaman}, \binits{F.}},
\bauthor{\bsnm{Guha}, \binits{B.K.}},
\bauthor{\bsnm{Ray}, \binits{S.}}:
\batitle{Anisotropic stars with non-static conformal symmetry}.
\bjtitle{Astrophys. Space Sci.}
\bvolume{361},
\bfpage{167}
(\byear{2016})
\end{barticle}
\endbibitem

\bibitem[\protect\citeauthoryear{Gokhroo and Mehra}{1994}]{GM1994}
\begin{barticle}
\bauthor{\bsnm{Gokhroo}, \binits{M.K.}},
\bauthor{\bsnm{Mehra}, \binits{A.L.}}:
\batitle{Anisotropic spheres with variable energy density in general relativity}.
\bjtitle{Gen. Relat. Gravit.}
\bvolume{26},
\bfpage{75}--\blpage{84}
(\byear{1994})
\end{barticle}
\endbibitem

\bibitem[\protect\citeauthoryear{Herrera}{1992}]{H1992}
\begin{barticle}
\bauthor{\bsnm{Herrera}, \binits{L.}}:
\batitle{Cracking of self-gravitating compact objects}.
\bjtitle{Phys. Lett. A}
\bvolume{165}(\bissue{3}),
\bfpage{206}--\blpage{210}
(\byear{1992})
\end{barticle}
\endbibitem

\bibitem[\protect\citeauthoryear{Abreu et~al.}{2007}]{AH2007}
\begin{barticle}
\bauthor{\bsnm{Abreu}, \binits{H.}},
\bauthor{\bsnm{Hernández}, \binits{H.}},
\bauthor{\bsnm{Núñez}, \binits{L.A.}}:
\batitle{Sound speeds, cracking and the stability of self-gravitating anisotropic compact objects}.
\bjtitle{Class. Quant. Gravit.}
\bvolume{24}(\bissue{18}),
\bfpage{4631}
(\byear{2007})
\end{barticle}
\endbibitem

\bibitem[\protect\citeauthoryear{Ponce~de Leon}{1993}]{P1993}
\begin{barticle}
\bauthor{\bsnm{Leon}, \binits{J.}}:
\batitle{Limiting configurations allowed by the energy conditions}.
\bjtitle{Gen. Relat. Gravit.}
\bvolume{25}(\bissue{11}),
\bfpage{1123}--\blpage{1137}
(\byear{1993})
\end{barticle}
\endbibitem

\bibitem[\protect\citeauthoryear{Visser}{1995}]{V1995}
\begin{bbook}
\bauthor{\bsnm{Visser}, \binits{M.}}:
\bbtitle{Lorentzian Wormholes: From Einstein to Hawking}.
\bpublisher{Springer},
\blocation{Berlin}
(\byear{1995})
\end{bbook}
\endbibitem

\bibitem[\protect\citeauthoryear{Maurya and Tello-Ortiz}{2019}]{MT2019}
\begin{barticle}
\bauthor{\bsnm{Maurya}, \binits{S.K.}},
\bauthor{\bsnm{Tello-Ortiz}, \binits{F.}}:
\batitle{Generalized relativistic anisotropic compact star models by gravitational decoupling}.
\bjtitle{Eur. Phys. J. C}
\bvolume{79},
\bfpage{85}
(\byear{2019})
\end{barticle}
\endbibitem

\bibitem[\protect\citeauthoryear{Chan et~al.}{1993}]{CH1993}
\begin{barticle}
\bauthor{\bsnm{Chan}, \binits{R.}},
\bauthor{\bsnm{Herrera}, \binits{L.}},
\bauthor{\bsnm{Santos}, \binits{N.O.}}:
\batitle{Dynamical instability for radiating anisotropic collapse}.
\bjtitle{Mon. Not. R. Astron. Soc.}
\bvolume{265}(\bissue{3}),
\bfpage{533}--\blpage{544}
(\byear{1993})
\end{barticle}
\endbibitem

\bibitem[\protect\citeauthoryear{Bondi}{1964}]{B1964}
\begin{barticle}
\bauthor{\bsnm{Bondi}, \binits{H.}}:
\batitle{The contraction of gravitating spheres}.
\bjtitle{Proc. R. Soc. Lond. A}
\bvolume{281}(\bissue{1384}),
\bfpage{39}--\blpage{48}
(\byear{1964})
\end{barticle}
\endbibitem

\bibitem[\protect\citeauthoryear{Moustakidis}{2017}]{M2017}
\begin{barticle}
\bauthor{\bsnm{Moustakidis}, \binits{C.C.}}:
\batitle{The stability of relativistic stars and the role of the adiabatic index}.
\bjtitle{Gen. Relat. Gravit.}
\bvolume{49}(\bissue{5}),
\bfpage{68}
(\byear{2017})
\end{barticle}
\endbibitem

\bibitem[\protect\citeauthoryear{Straumann}{1984}]{S1984}
\begin{bbook}
\bauthor{\bsnm{Straumann}, \binits{N.}}:
\bbtitle{General Relativity and Relativistic Astrophysics},
p. \bfpage{43}.
\bpublisher{Springer},
\blocation{Berlin}
(\byear{1984})
\end{bbook}
\endbibitem

\bibitem[\protect\citeauthoryear{Buchdahl}{1959}]{B1959}
\begin{barticle}
\bauthor{\bsnm{Buchdahl}, \binits{H.A.}}:
\batitle{General relativistic fluid spheres}.
\bjtitle{Phys. Rev.}
\bvolume{116},
\bfpage{1027}--\blpage{1034}
(\byear{1959})
\end{barticle}
\endbibitem

\bibitem[\protect\citeauthoryear{Bowers and Liang}{1974}]{BL1974}
\begin{barticle}
\bauthor{\bsnm{Bowers}, \binits{R.L.}},
\bauthor{\bsnm{Liang}, \binits{E.P.T.}}:
\batitle{Anisotropic spheres in general relativity}.
\bjtitle{Astrophys. J.}
\bvolume{188},
\bfpage{657}--\blpage{665}
(\byear{1974})
\end{barticle}
\endbibitem

\bibitem[\protect\citeauthoryear{Barraco et~al.}{2003}]{BH2003}
\begin{barticle}
\bauthor{\bsnm{Barraco}, \binits{D.E.}},
\bauthor{\bsnm{Hamity}, \binits{V.H.}},
\bauthor{\bsnm{Gleiser}, \binits{R.J.}}:
\batitle{Anisotropic spheres in general relativity reexamined}.
\bjtitle{Phys. Rev. D}
\bvolume{67},
\bfpage{064003}
(\byear{2003})
\end{barticle}
\endbibitem

\bibitem[\protect\citeauthoryear{Böhmer and Harko}{2006}]{BH2006}
\begin{barticle}
\bauthor{\bsnm{Böhmer}, \binits{C.G.}},
\bauthor{\bsnm{Harko}, \binits{T.}}:
\batitle{Bounds on the basic physical parameters for anisotropic compact general relativistic objects}.
\bjtitle{Class. Quant. Gravit.}
\bvolume{23}(\bissue{22}),
\bfpage{6479}
(\byear{2006})
\end{barticle}
\endbibitem

\end{thebibliography}

\end{document}